\documentclass[doublespacing]{elsart}
\usepackage{amssymb}

\begin{document}

\begin{frontmatter}

\title{Electron-impact broadening of the $3s-3p$ lines in low-Z Li-like ions}
\author[We]{Yu. V. Ralchenko\corauthref{cor}},
\corauth[cor]{Corresponding author. Present
address: Atomic Physics Division, National Institute of
Standards and Technology, Gaithersburg, MD 20899-8422}
\ead{yuri.ralchenko@nist.gov}
\author[Ma]{H. R. Griem},
\author[Au]{I. Bray}
\address[We]{Faculty of Physics, Weizmann Institute of Science, Rehovot 76100,
Israel}
\address[Ma]{Institute for Research in Electronics and Applied Physics,
University of Maryland, College Park MD 20742}
\address[Au]{Centre for Atomic, Molecular and Surface Physics, School of
Mathematical and Physical Sciences, Murdoch University, Perth 6150, Australia}


\begin{abstract}
The collisional electron-impact line widths of the $3s-3p$ transitions in
Li-like ions from B III to Ne VIII are calculated with the convergent
close-coupling (CCC) method from the atomic collision theory. The elastic
and inelastic contributions to the line broadening and their Z-scaling are
discussed in detail, and comparisons with  recent experimental and
theoretical results are also presented. It is found that similar to our
previous study of line broadening in Be-like ions, the difference
between experimental and CCC results monotonically increases with the
spectroscopic charge of an ion.
\end{abstract}

\end{frontmatter}

\section{Introduction}

The broadening of spectral lines in dense plasmas primarily stems from a
complex interaction of an emitting atom with the perturbing plasma
constituents \cite{Griem74}. Accordingly, the line shapes and shifts reveal
(generally non-trivial) dependence on major plasma characteristics such as
particle density and temperature, and hence can serve as important
diagnostic tools. The complexity of the interaction, however, often impedes
a reliable understanding of the line broadening, both experimentally and
theoretically.

Although the line widths in hydrogen and hydrogen-like ions can normally be
explained  to a high level of accuracy, the recent studies on collisional
broadening of isolated lines of multiply-charged ions have unveiled
worrying disagreement between the experimental data and various 
theoretical calculations, including the advanced quantum-mechanical 
ones . An example is
provided by the resonance $2s-2p$ line in the B III ion for which the
experimental width \cite{Glenzer96} was found to be about a factor of two
larger than non-perturbative results of both quantal and semiclassical \cite
{Griem1997,AlexLee}. As for the $2s3s-2s3p$ singlet and triplet lines in
Be-like ions from boron to neon, it was shown that although the quantal
results agree well with experimental line widths for low-charge ions, the
deviation between experiment and theory monotonically increases along the
isoelectronic sequence \cite{Ral1999,JQSRT01}, reaching  a factor of two
for Ne VII. Such  disagreement is indeed disturbing since the collisional
broadening of the isolated lines in multiply-charged ions is primarily
determined by the inelastic binary collisions, which in turn can be either
experimentally assessed (as was done for the B III $2s-2p$ excitation cross
section \cite{Woitke}) or  reliably cross-checked by comparing with
other theoretical calculations.

A reliable measurement of rather narrow line widths in multiply-charged ions
poses demanding requirements both on the generation of the plasma 
properties required and the methods for
an {\em independent} determination of electron density $N_{e}$ and temperature
$T_{e}$. In such benchmark experiments the laboratory plasmas, or plasma
volumes, that are used in line broadening measurements have to be as uniform
and stationary as possible and, if at all possible, to be free of various
instabilities and turbulences. Needless to say, these requirements are
not easily met in laboratory plasmas and are difficult to verify in 
astrophysical plasmas. Moreover, an
independent measurement of $N_{e}$ and $T_{e}$ requires  the most
accurate available techniques which may not be applicable in some plasma
conditions. It is thus not surprising that systematic measurements of
broadening of isolated lines in multiply-charged ions were primarily 
conducted by
two groups, namely, at the Institute of Physics in Belgrade, 
Yugoslavia, and at the Ruhr Universit\"{a}t in Bochum, Germany. Both 
groups have recently reported
measurements of the $3s-3p$ line widths in Li-like ions from boron to oxygen
(using a linear arc  \cite{Blago99}) and from carbon to neon (using a gas-puff
Z-pinch \cite {Glenzer92,Glenzer93,Hegazy}), respectively, and the major plasma
characteristics were independently measured for each shot. Furthermore, several
other experimental results as well as a number of theoretical calculations are
available for these transitions. Thus, the existing data could when 
taken as a whole provide a diverse set of results for comparison of 
measurements and theory.

In continuation of the previous quantal calculations of line broadening in
multiply-charged ions \cite{Griem1997,Ral1999,JQSRT01}, we present here a
quantum-mechanical calculation of the electron-impact line widths of 
the $3s-3p$
transitions in Li-like ions from B III to Ne VIII. The paper is organized as
follows. Section II contains the basic discussion of the theoretical method
used for the present calculation. The convergent close-coupling method is
applied for determination of both inelastic and elastic contributions to the
line widths. The results of calculations as well as the comparison with
experimental data are presented in Sec. III. The following Section deals with
the scaling properties of the line widths and finally the conclusions are drawn
in the last Section.

\section{Theory}

The calculational method implemented in the present work has been described in
detail previously (see, e.g., Ref. \cite{JQSRT01} and references therein), and
hence only the principal steps in the line width calculation are outlined here.
The electron-impact full width at half-maximum (FWHM) of an isolated spectral
line is calculated here using the Baranger formula \cite{Bar583}, which
represents the line width as a product of the electron density $N_{e}$ and the
sum of the thermally-averaged (i) inelastic (excitation, deexcitation,
ionization, etc.) cross sections for all possible collisional transitions from
the upper and lower levels of the radiative transition in question, 
and (ii) the
difference squared of the non-Coulomb elastic amplitudes $f_{u}$ and $f_{l}$ of
scattering from the upper and lower levels. This representation allows one to
efficiently apply the computational methods developed for a general atomic
collision theory to calculations of the impact line widths.

In this work both inelastic cross sections and elastic amplitudes are
calculated with the convergent close-coupling (CCC) method \cite{Bray2},
which is presently considered to be one of the most accurate techniques for
electron-ion scattering problems. A detailed description of the method as
well as recent applications of the CCC technique can be found in the review
of Bray et al \cite{Bray2002} and references therein. We only mention here
that the basic idea of this method is the discretization of the continuum
along with the use of the square-integrable Laguerre basis, which
significantly facilitates the computational efforts.

It is well known that in multiply-charged ions the major contribution to the
line width originates from excitation and deexcitation, and therefore 
a reliable
estimate of the cross section accuracy would be important for an assessment of
the line width accuracy. Since the existing experimental techniques 
do not allow
one to measure the cross sections for transitions between excited states due to
their small lifetimes, a comparison of different {\em theoretical} results
seems presently to be the only tool for the error estimates. Fortunately, the
Li-like ions represent a simple quasi-one-electron atomic system with
practically pure LS-coupling for the lowest excited states (including the $n=3$
terms), so that one would expect the modern calculational techniques to produce
fairly accurate results for the collisional cross sections and amplitudes.
Indeed, it was confirmed in a number of publications that the CCC excitation
cross sections agree well with both the experimental data and other theoretical
non-perturbative results for the ground state excitations. Such comparisons
were done, e.g., for the R-matrix with pseudostates and K-matrix method results
\cite {Bar97,March97,Starob2002}. For higher-Z ions ($Z\geq 4$), a detailed
comparison of all excitation cross sections between the states with the
principal quantum number $n\leq 4$ has shown that the CCC data practically
coincide with the perturbative Coulomb-Born-exchange calculations \cite
{Liions}. Furthermore, a recent paper on electron-impact excitation in Be II
and B III (also including the transitions between the excited states up to
$n\leq 4$) again demonstrated a high level of agreement between the CCC\ and
the K-matrix cross sections \cite{Starob2002}. Thus, the available data
comparisons indicate that the CCC method seems to provide high accuracy for the
excitation cross sections of Li-like ions and, correspondingly, for the
inelastic contributions to the collisional line broadening. Also, it has to
be added that, as one would expect, for all ions considered here the major
contribution to the inelastic part comes from the collisional dipole-allowed
$3s \rightarrow 3p$, $3p \rightarrow 3s$\ and  $3p\rightarrow 3d$\ channels
that amount to about 90\% of the inelastic line width. The contribution of
ionization and recombination channels was found to be negligible compared to
(de)excitation processes.

The accuracy of the theoretical elastic contribution to the line widths still
remains largely unknown. The main reason is that the elastic non-Coulomb
scattering amplitudes as well as the angular-integrated elastic difference term
(EDT) $\sigma _{EDT}\left( E\right) \equiv \int \left| f_{u}\left( \theta%
\right) -f_{l}\left( \theta \right) \right| ^{2}d\Omega $ are practically
unavailable in the literature. However, given the high accuracy of the CCC
inelastic cross sections, one would expect about the same accuracy for the
elastic part as well. The EDT's for all ions from B III to Ne VIII have been
calculated in the present work (see the Z-scaled results in Sect. 4). It was
found that while the elastic cross sections $\sigma_{s,p}\left( E\right)%
\equiv \int \left| f_{s,p}\left( \theta \right) \right| ^{2}d\Omega $ follow
the $1/E$ law for practically all energies from 0.1 to 100 eV, the EDTs show a
sharper fall $\sigma _{EDT}\left( E\right) \sim 1/E^{\alpha }$ with $\alpha%
=1.35\div 1.45$, thereby pointing out a strong cancellation in the difference
of elastic amplitudes. This steeper dependence on electron energy had already
been noticed in our previous studies on the B III $2s-2p$ line and Be-like
$2s3s-2s3p$ transitions \cite{Griem1997,JQSRT01}.

\section{Results and comparisons}

The calculated elastic, inelastic and total FWHMs (in units of \AA ) for the
electron density of 10$^{18}$ cm$^{-3}$ vs. electron temperature $T_{e}$ in the
range of $T_{e}$ = $2-20$ eV are presented in Table I for the ions from B III
to F VII and in the range of $T_{e}$ = $2-50$ eV for Ne VIII, where
experimental data are available for higher $T_{e}$. The inelastic contribution
is seen to dominate over the elastic part for all calculated temperatures,
although for small $T_{e}$ = 2 eV the relative contribution of the elastic
line width reaches as much as 50\% for N V and O VI.The $T_{e}$-scaling of the
calculated line widths for $T_{e}$ = $2-20$ eV is approximately given as

\begin{equation}
\lambda \propto T_{e}^{-\alpha },
\end{equation}

where $\alpha$ takes values of 0.17 (B III), 0.29 (C IV), 0.28 (N V), 0.31 for
O VI, 0.26 (F VII) and 0.23 (Ne VIII). These values show a weaker than the
1/$T_{e}^{0.5}$ dependence which is often used in the literature.

Figures 1-6 present the calculated CCC line widths and the available
experimental and theoretical data scaled to the electron density of N$_{e}$ =
10$^{18}$ cm$^{-3}$. Most of the other theoretical results are taken from Ref.
\cite{Blago99}. Below we often make use of the following designations: DSB --
semiclassical results by Dimitrijevi\'{c} and Sahal-Brechot as cited in Ref.
\cite{Blago99}, DK -- calculations made within the modified semiempirical
method by Dimitrijevi\'{c} and Konjevi\'{c} as cited in Ref. \cite{Blago99}, HB
-- semiclassical results by Hey and Breger \cite{HeyBre82}, MNPSC -- modified
non-perturbative semiclassical method by Alexiou \cite{Alexiou}.

Note also that the measurements by the Belgrade group were carried out for low
densities N$_{e}=(0.3\div 1.4)\times 10^{17}$ cm$^{-3}$, while the
Bochum group measurements were performed on a linear gas-puff Z-pinch at
densities exceeding $10^{18}$ cm$^{-3}$. The experimental error bars shown on
the plots do not include the uncertainty of electron density measurements which
can be as high as 15\%.  We now discuss the calculations for 
individual ions in more detail.

\subsection{B III (Fig. 1)}

For this ion, the $3s-3p$ line width was measured only by the Belgrade group
\cite{Blago99}. The typical experimental electron densities and temperatures
were about 10$^{17}$ cm$^{-3}$ and 5 eV, respectively. This is the only case
where the present CCC results practically coincide with the experimental data.
Similarly, for the Be-like ions \cite{JQSRT01} an agreement was found only for
the lowest-charge ion of B II. The semiclassical results of Griem and
DSB, as cited in Ref. \cite{Blago99},
exceed the experimental data by about 70 and 40\%, respectively, while the
semiempirical data of DK are seen to be
within the errors bars, exceeding the CCC line widths by 15-20\% over the
temperature interval $T_{e}$ = $2-20$ eV. The only other quantum-mechanical
results, by Seaton \cite{Seaton88}, also confirm the experimental data near
$T_{e}$ = $5.5-6$ eV and, moreover, agree with the present 
calculations to within
10\% over a wider range of T$_{e}$. Yet such good agreement for low
temperatures may not be a measure of good agreement of elastic parts since the
present calculations show that even for T$_{e}$ as low as 2 eV the elastic
contribution is only about one-fifth of the total line width (see Table I).

\subsection{C IV (Fig. 2)}

The theoretical and experimental data for the C IV $3s-3p$ line width are quite
extensive. The older measurements by Bogen \cite{Bogen} and by El-Farra and
Hughes \cite{EFH} are well confirmed by the recent results of the Belgrade
group \cite{Blago99}. The gas-puff Z-pinch data from Ref. \cite{Glenzer92},
although being somewhat higher than the extrapolated $T_{e}$-trend from other
measurements, nevertheless overlap with the error bars from the Belgrade
results. The recent measurements of Sreckovi\'{c} et al. 
\cite{Sreckovic00} on a
linear low-pressure pulsed arc were performed for low temperatures 
$T_{e} \sim 2$ eV
and showed a significant spread. As for the theoretical results, 
Seaton \cite {Seaton88} carried out a quantum-mechanical R-matrix 
calculation within the
framework of the Opacity Project for this line width taking into account
the perturbing $n=2$ and $n=3$ states. Later Burke \cite{Burke92} improved
Seaton's calculations using the same method but adding $n=4$ states, which
resulted in a moderate increase of less than 10\% to the line widths. The
present CCC results are seen to agree excellently with Burke's calculations,
particularly for higher temperatures where the inelastic contribution 
dominates.
Even for the lowest temperatures ($T_{e}\sim 4$ eV) the difference is
well within 10\%. However, as can be seen from Fig. 2, the recent experimental
results \cite {Blago99,Glenzer92} exceed the CCC line widths by up to 30\%.

\subsection{N V (Fig. 3)}

Experimental line widths for N V are available from both Bochum and Belgrade
experiments, the former having rather large error bars for the electron
temperature. Of the available theoretical results, the semiclassical method of
Griem \cite{Griem74} provides very good agreement with both sets of
experimental data over the range of temperatures from 6
to 24 eV. It is noted that the $T_{e}-$dependence of line widths
is very similar for the CCC, Griem and DK results, while the semiclassical DSB
data seem to decay faster for higher $T_{e}$.

\subsection{O VI (Fig. 4)}

The available measurements for this ion were carried out over a rather large
range of electron temperatures, $T_{e}$ = $5-18$ eV, although the error bars
here are quite significant, and therefore it is difficult to reach
conclusions regarding the temperature dependence of the data. Figure 4 shows
that both sets of experimental data for O VI well agree with the semiclassical
calculations of DSB. The only other quantum-mechanical data of Seaton for
$T_{e}$ $\approx $ $11$ eV practically coincide with the present calculations,
and the DK results are again within 20\% from the CCC data. The semiclassical
results by Griem, Hey and Breger, and Alexiou agree with each other and are at
the lower edge of the experimental error bars.

\subsection{F VII (Fig. 5)}

The only available experimental data for this line for $T_{e}=14-18$ eV and
electron densities in the range $(1.5-3)\times 10^{18}$ cm$^{-3}${\em \ } were
reported in Ref. \cite{Glenzer93}. Again, similar to the O VI case, the
agreement with the semiclassical results of DSB is very good, while all other
theoretical results are much lower than the experiment.

\subsection{Ne VIII (Fig. 6)}

This is the case where the difference between the experiment and {\em all}
existing theoretical results is the largest. The broadening of the Ne VIII
$3s-3p$ line at the highest electron temperatures of 30 and 42 eV was measured
by the Bochum group a decade ago \cite{Glenzer92,Glenzer93} and has recently
been remeasured on the same experimental setup \cite{Hegazy}. The 
latest data are
seen to be closer to the theoretical values (Fig. 6), so that DSB and Griem's
results are at the lower edge of the error bars. However, the other available
theoretical data are much lower, and the disagreement with the experiment
exceeds 50\% for the MNPSC method, DK and HB results, while the CCC line widths
are factor 2.3 smaller than the data from Ref. \cite{Hegazy}.

\section{Discussion and Scaling Properties}

As has already been mentioned in the Introduction, in the previous work on
collisional broadening in Be-like ions \cite{JQSRT01} we found that the CCC
results and experimental data systematically diverge along the isoelectronic
sequence. A reader might have already noticed that this is also the case in the
present comparison. Figure 7 shows the ratio of experimental to calculated CCC
line widths for the $3s-3p$ transition as a function of atomic number for all
Belgrade and Bochum data points presented in Figs. 1-6. (The two older data
points for Ne VIII \cite{Glenzer92} are not included in this plot.) The ratios
using the Belgrade experimental results are shown by the large shaded
circles, while those using the Bochum data are shown by the large open circles.
The dashed line is added to the plot in order to make the trend more visible.
Further, we add the ratios for the $2s3s-2s3p$ line widths in Be-like
ions \cite{JQSRT01} (small shaded and open squares for the Bochum and
Belgrade data, respectively) to emphasize that the ratios are indeed very close
for both isoelectronic sequences.

The fact that a very similar behavior in terms of the ratio between the
experimental and quantum-mechanical line widths is found for completely
different experiments suggests a similar physical rather than 
instrumental or systematic experimental
mechanism responsible for the deviation of experimental and theoretical
widths.   As has already been mentioned above, in most of the experiments an
independent diagnostic is required for a reliable determination of
the electron density (and temperature). In order to eliminate the broadening
due to the opacity effects, the measurements were done using small amounts of
test gas containing the impurity ions added to a major bulk gas. In both series
of experiments, $N_{e}$ was determined from the bulk plasma (hydrogen in Bochum
experiments and helium in Belgrade measurements) using 90$^{o}$ Thomson
scattering of laser light, or a well-known dependence of the Paschen-$\alpha $
line of He II on $N_{e}$, respectively. The resolved impurity peak in the
former measurements indicated that the impurity density was $\leq $
1\% of $N_{e}$. Such measurements are obviously carried out macroscopically and
thus are insensitive to the density fluctuations near the multi-charged
impurity ions. The local increase of electron density near an impurity ion
which has a higher Coulomb charge than the background ions is obvious already
from the Debye-H\"{u}ckel picture of a charge screening in plasmas. The recent
detailed classical many-body and molecular-dynamics calculations \cite{Talin01}
of non-linear behavior of electrons near a positive impurity ion for various
plasma conditions (with a strong coupling constant $\Gamma =0.03-0.5$ which is
much larger than that in both Belgrade and Bochum experiments) indeed show an
increase of electron  density in the ion vicinity. However, preliminary
estimates based on the Debye-H\"{u}ckel picture do not seem to support the
importance of this phenomenon for line broadening in the experiments discussed
here. Needless to say, a detailed
investigation of how the local density fluctuations near impurity ions could
influence the spectral line broadening would obviously be very helpful.

It should also be mentioned that the extra contribution to the line widths
from ion-impact broadening may be quite reliably estimated using the
semiclassical calculations of Ref. \cite{Blago99}. For the ions from B III
to O VI the ion width was found to not exceed 5\% of the electron FWHM and
thus it can be safely neglected here.

The scaling of the impact widths for isolated lines has long been a subject of
detailed studies. The possibility of accurately predicting the 
unknown line widths
from those already measured and/or calculated  is always a strong impetus to
attack this problem. The regularities of the Stark broadening along
isoelectronic sequences have been discussed for some time in connection with
the critical reviews of Stark broadening data (see, e.g., \cite{review} and
references therein). The scaling of the line widths with the ion spectroscopic
charge can be easily derived from the Baranger formula. If one assumes that
only one perturbing transition contributes to the line width, then 
the inelastic
contribution is proportional to:

\begin{equation}
\Delta \lambda _{inel}\propto \lambda ^{2}\left\langle \sigma
v\right\rangle .
\end{equation}

For the $\Delta n=0$ transition the wavelength scales as $\lambda \propto 1/Z%
$, the threshold cross section $\sigma \propto 1/Z^{3}$, so that for the
line width one obtains $\Delta \lambda _{inel}\propto 1/Z^{5}$ for the same
electron temperatures. Our previous study of excitation cross sections in
Li-like ions with spectroscopic charge $Z\geq 4$ \cite{Liions} showed that the
scaling of the CCC cross sections slightly differs from the $1/Z^{3}$
dependence, and for the $3s-3p$ and $3p-3d$ transitions, which are dominant
in the present case, the scaling is $1/\left( Z+1.75\right)^{3.4}$
and $1/\left( Z+0.44\right) ^{3.5}$, respectively. This results in modified
scaling properties for the CCC line widths which are found to be approximately

\begin{equation}
\Delta \lambda _{inel}^{CCC}\propto 1/Z^{17/4}
\end{equation}

for the ions considered in this work. The scaled products $\Delta \lambda%
_{inel}^{CCC}\times Z^{17/4}$ for all ions from B III to Ne VIII are presented
in Fig 8(a). One can see that the scaling is accurate to only 10\% for
all values of $T_{e}$ and significantly improves with the increase of the
spectroscopic charge.

The scaling of the elastic part of the line width, however, is not as obvious
as that of the inelastic one. For the problem in question the elastic
scattering is dominated by the polarization potential $\sim 1/r^{4}$
resulting from interactions between the $n=3$ states. The solutions for the
elastic cross sections in a central field of the modified Blumington
polarization potential \cite{McD}

\begin{equation} V\left( r\right) =-\frac{\alpha _{D}}{2\left(
r^{2}+d^{2}\right) ^{2}}, \end{equation}

where $\alpha _{D}$ is the dipole polarizability and $d$ is the characteristic
length of the order of the atomic orbit, are well known for the Born and
semiclassical approximations. Since for the $\Delta%
n=0$ transitions $\alpha _{D}\sim 1/Z^{3}$ and assuming $d\sim 1/Z,$ the
Z-scaling in these cases is $\sigma \sim 1/Z^{3}$ and $\sigma \sim 1/Z^{7/3}$,
respectively. However, we find that the elastic cross sections $3s-3s$ and
$3p-3p$ as well as the elastic difference term exhibit a weaker scaling law,
namely, $\sigma \sim 1/Z^{3/2}$. This difference may result from the fact that
the theoretical $Z$-dependence of the Born and semiclassical cross sections is
ususally derived assuming that the major contribution comes from the high
partial waves $L\gg 1$. On the contrary, the present CCC calculations show
that most of the elastic cross sections is provided by the low
partial waves with $L\lesssim 3-4$ only. Returning to the Z-scaling of the
elastic line widths, one immediately obtains that for the same electron
temperatures and $\sigma _{EDT}\sim 1/Z^{3/2}$ the elastic contribution to the
line width scales as

\begin{equation}
\Delta \lambda _{el}^{CCC}\propto 1/Z^{7/2}.
\end{equation}

Somewhat better accuracy is actually achieved with a $1/Z^{15/4}$ scaling
which is presented in Fig. 8(b), where the scaled line widths $\Delta%
\lambda_{el}^{CCC}\times Z^{15/4}$ are plotted vs. electron temperature.

\section{Conclusions}

One of the most powerful techniques in the atomic collision theory--the
convergent close-coupling method--is applied here to the electron-impact
broadening of the isolated $3s-3p$ lines in Li-like ions from B III to Ne VIII.
The elastic and inelastic contributions to the line width are explicitly
calculated and tabulated for electron temperatures from 2 to 20 eV. A
comparison with available experimental data (mostly from the Belgrade and
Bochum groups) shows that although the quantum-mechanical data presented here
agree well with the measurements for lowest-charge ions, the disagreement
progressively increases with an increase of the ion charge. In no case is the
experimental line width {\em smaller} than the quantum-mechanical one,
suggesting some additional line broadening process in the higher-Z experiments.

\section{Acknowledgments}

We are grateful to H.-J. Kunze and Y. Maron for valuable discussions and
comments and to S. Alexiou for providing the MNPSC data prior to publication.
This work was supported in part by the Israeli Ministry of Absorption and
Israeli Academy of Sciences (Yu. V. R.) and by the US National Science
Foundation (H.R.G.). I.B. acknowledges the support of the Australian Research
Council.

\newpage

Figure captions.

Figure 1. Collisional line widths for the 3s-3p transition in B III. 
Experiment:
Ref. \cite{Blago99} --$ \Box$. Theory: present work -- solid line with solid
circles, Griem (as cited in Ref. \cite{Blago99}) -- dot line, DSB
\cite{Blago99} -- short-dash line, Seaton \cite{Seaton88} -- dot-dash line,
DK (as cited in Ref. \cite{Blago99}) -- long-dash line.

Figure 2. Collisional line widths for the 3s-3p transition in C IV. Most of
notations are the same as in Fig. 1. Experiment:  Ref. \cite{Glenzer92} --
$\bigcirc$, Ref. \cite{Bogen} -- $\Diamond$, Ref. \cite{EFH} --
$\bigtriangleup$, Ref. \cite{Sreckovic00} -- $\bigtriangledown$. Theory:
Alexiou \cite{Alexiou} -- $\times$, Burke \cite{Burke92} -- dot-double-dash
line.

Figure 3. Collisional line widths for the 3s-3p transition in N V. Most of
notations are the same as in Fig. 1 and 2.

Figure 4. Collisional line widths for the 3s-3p transition in O VI. Most of
notations are the same as in Fig. 1 and 2. Theory: Seaton \cite{Seaton88} --
$\ast$, Hey and Breger \cite{HeyBre82} -- dash-double-dot line.

Figure 5. Collisional line widths for the 3s-3p transition in F VII. Most of
notations are the same as in Fig. 4. Experiment:  Ref. \cite{Glenzer93} --
$\bigcirc$.

Figure 6. Collisional line widths for the 3s-3p transition in Ne VIII. Most of
notations are the same as in Fig. 4. Experiment: Ref. \cite{Hegazy} -- shaded
circle.

Figure 7. Ratio of experimental and quantum-mechanical line widths vs. atomic
number. Li-like ions: circles, Be-like ions: squares. Ratios with the Belgrade
data correspond to shadowed symbols, while ratios with the Bochum data
correspond to open symbols.

Figure 8. Scaled (a) inelastic and (b) elastic CCC line widths. B III -- solid
line, C IV -- short-dash line, N V -- dotted line, O VI -- dot-dashed line, F
VII -- long-dash line, Ne VIII -- dot-double-dash line.

\newpage

Table I. Electron-impact linewidths (in \AA ) for the $3s-3p$ transition in
B III -- Ne VIII ions. The electron temperature T$_{e}$ is in eV and
the electron density is 10$^{18}$ cm$^{-3}$.

\medskip

\begin{tabular}{cccccccccccccc}
\hline
T$_{e}$ &  &  & B III &  &  &  & C IV &  &  &  & N V &  \\
\cline{3-5}\cline{7-9}\cline{11-13}
&  & el & inel & total &  & el & inel & total &  & el & inel & total & \\
\hline
2 &  & 2.860 & 10.25 & 13.11 &  & 1.117 & 3.799 & 4.916 &  & 0.661 & 1.348 &
2.009 &  \\
4 &  & 1.578 & 9.880 & 11.46 &  & 0.589 & 3.490 & 4.079 &  & 0.342 & 1.318 &
1.656 &  \\
6 &  & 1.096 & 9.576 & 10.67 &  & 0.405 & 3.212 & 3.617 &  & 0.229 & 1.254 &
1.483 &  \\
8 &  & 0.844 & 9.344 & 10.19 &  & 0.312 & 3.007 & 3.319 &  & 0.172 & 1.197 &
1.369 &  \\
10 &  & 0.689 & 9.163 & 9.852 &  & 0.255 & 2.852 & 3.107 &  & 0.138 & 1.151
& 1.289 &  \\
12 &  & 0.584 & 9.007 & 9.591 &  & 0.217 & 2.729 & 2.946 &  & 0.115 & 1.111
& 1.226 &  \\
14 &  & 0.508 & 8.878 & 9.386 &  & 0.189 & 2.629 & 2.818 &  & 0.0981 & 1.076
& 1.174 &  \\
16 &  & 0.450 & 8.747 & 9.197 &  & 0.168 & 2.544 & 2.712 &  & 0.0857 & 1.047
& 1.133 &  \\
18 &  & 0.404 & 8.659 & 9.063 &  & 0.151 & 2.472 & 2.623 &  & 0.0761 & 1.021
& 1.097 &  \\
20 &  & 0.367 & 8.531 & 8.898 &  & 0.137 & 2.410 & 2.547 &  & 0.0684 & 0.996
& 1.064 &  \\ \hline
\end{tabular}

\bigskip \newpage

Table I (continued).
\medskip

\begin{tabular}{cccccccccccccc}
\hline
T$_{e}$ &  &  & O VI &  &  &  & F VII &  &  &  & Ne VIII &  \\
\cline{3-5}\cline{7-9}\cline{11-13}
&  & el & inel & total &  & el & inel & total &  & el & inel & total 
& \\ \hline
2 &  & 0.340 & 0.668 & 1.008 & &  0.141 & 0.334 & 0.475 &  & 0.0732 & 0.177
& 0.250 \\
4 &  & 0.178 & 0.664 & 0.842 & &  0.0712 & 0.342 & 0.413 &  & 0.0413 & 0.182
& 0.223 \\
6 &  & 0.118 & 0.624 & 0.743 & &  0.0490 & 0.327 & 0.376 &  & 0.0289 & 0.176
& 0.205 \\
8 &  & 0.0881 & 0.588 & 0.676 & &  0.0382 & 0.310 & 0.348 &  & 0.0226 & 0.169
& 0.192 \\
10 &  & 0.0701 & 0.558 & 0.629 & &  0.0318 & 0.295 & 0.327 &  & 0.0187 & 0.162
& 0.180 \\
12 &  & 0.0581 & 0.533 & 0.591 & &  0.0274 & 0.282 & 0.310 &  & 0.0161 & 0.156
& 0.172 \\
14 &  & 0.0496 & 0.513 & 0.562 & &  0.0242 & 0.271 & 0.296 &  & 0.0141 & 0.150
& 0.165 \\
16 &  & 0.0433 & 0.496 & 0.539 & &  0.0217 & 0.262 & 0.284 &  & 0.0126 & 0.146
& 0.159 \\
18 &  & 0.0384 & 0.481 & 0.519 & &  0.0197 & 0.254 & 0.273 &  & 0.0114 & 0.143
& 0.154 \\
20 &  & 0.0345 & 0.468 & 0.503 & &  0.0180 & 0.247 & 0.265 &  & 0.0104 & 0.139
& 0.150 \\
30 &  &  &  &  &  &  &  &  &  & 0.0073 & 0.128 & 0.135 \\
40 &  &  &  &  &  &  &  &  &  & 0.0055 & 0.120 & 0.125 \\
50 &  &  &  &  &  &  &  &  &  & 0.0044 & 0.114 & 0.119 \\ \hline
\end{tabular}

\end{document}